%% file: paper.tex
\documentclass[times, 10pt,twocolumn]{article} 
\usepackage{latex8}
\usepackage[small,compact]{titlesec}
\usepackage{times}
\usepackage{graphicx}
\usepackage{subfigure}
\usepackage{algorithmic}
\usepackage{algorithm}
\usepackage{amsthm}
\usepackage{setspace}
\usepackage[hmargin={0.75in,0.75in},vmargin={0.75in,0.85in}]{geometry}

\begin{document}

\twocolumn[% 
\centerline{\LARGE \bf \textsf{Overlay Structure for Large Scale
 Content Sharing:}}
\centerline{\LARGE \bf \textsf{Leveraging Geography as the Basis for Routing
  Locality}} %% Paper title
 \medskip
 \centerline{\large {\bf Shah Asaduzzaman} and {\bf Gregor v. Bochmann}}
\centerline{\large SITE, University of Ottawa, Ottawa, ON, Canada K1N 6N5}
\centerline{\texttt {\large \{asad,bochmann\}@site.uottawa.ca}}
      %% Author name
 \bigskip
 ]

%% \title{Overlay Structure for Large Scale Content Sharing:\\ Leveraging
%%   Geography as the Basis for Routing Locality}
%% %\title{Geography as the Basis of Generalized Overlay Routing Service}

%% \author{Shah Asaduzzaman and Gregor v. Bochmann \\
%% School of Information Technology and Engineering\\
%% University of Ottawa, Ottawa, ON, Canada K1N 6N5\\
%% \texttt{\{asad,bochmann\}@site.uottawa.ca}
%% }

%% \maketitle
%\thispagestyle{empty}

\input{abstract}
\input{intro}

\input{separation} % need for separation of search and transport
\input{objectives} % design objectives of the tansport
		   % overlay
\input{geography} % review of topology aware overlays and reasoning
                  % for use of geography

\input{structure} % review of different overlay structures, that are
                  % independent of the coordinate space and
                  % introducing oure hierarchical space partitioning
                  % based overlay
%\input{analysis}
\input{app}   % discussion on how different applications can take
                  % benefit of the geographic overlay
\input{conclusion}

%------------------------------------------------------------------------- 

\bibliographystyle{latex8}
\bibliography{geoDHT}

\end{document}

%% file: abstract.tex
%%% Local Variables: 
%%% mode: latex
%%% TeX-master: "paper"
%%% End: 
\begin{abstract}
  In this paper we place our arguments on two related issues in the
  design of generalized structured peer-to-peer overlays. First, we
  argue that for the large-scale content-sharing applications, lookup
  and content transport functions need to be treated
  separately. Second, to create a location-based routing overlay
  suitable for content sharing and other applications, we argue that
  off-the-shelf geographic coordinates of Internet-connected hosts can
  be used as a basis. We then outline the design principles and
  present a design for the generalized routing overlay based on
  adaptive hierarchical partitioning of the geographical space.
\end{abstract}

%% file: intro.tex
%%% Local Variables: 
%%% mode: latex
%%% TeX-master: "paper"
%%% End: 
\section{Introduction}
Peer-to-peer overlay networks are nowadays envisioned as a single
self-organized and decentralized substrate to be used by many
different large-scale networked applications. Since inception of the
concept, a large number of alternative designs for peer-to-peer
overlays have been proposed to support different application needs.
At some point, consensus has emerged that there is a need for a single
well-defined interface to encapsulate the generic features provided by
these overlays. Progress have been made in this direction and several
well-understood features have been
defined~\cite{CommonAPI2003,OpenDHT2005}.

Because the evolution of peer-to-peer research has historically
depended on the content sharing applications, two fundamental
functional requirements of these applications -- search and transport
have guided the design of general purpose overlays. Initially, the
problem of search or lookup dominated the research while the transport
related problems gradually emerged. Researchers then have attempted to
accommodate transport related functionalities in generalized versions
of the overlays that were primarily created for
lookup~\cite{Scribe2002}. Also, randomness was introduced in the
overlay structures for several puposes such as load balancing and
source anonymity~\cite{Pastry2001,Chord2001}. As a result, the
conventional wisdom of routing locality in bulk data transport
necessary for efficient usage of network resources has been ignored.

With the rising of complaints from the ISPs against the peer-to-peer
traffic, there have been several proposals for introducing locality
awareness in the overlay structures~\cite{mOverlay2004, Le2005,
  Pietzuch2006, Xu2003, Mithos2003}. In most of the cases, localities
are defined based on explicit measurements of some application level
metric such as latency. This class of overlays, denoted as network
aware overlays suffer from the large background overhead of the
measurement.

In this paper, we argue that the geographic location of the end-hosts,
at the available granularity of ISP's points of presence, can be used
as the basis of a locality-based routing overlay. The argument is
founded on two observations. First, the Internet infrastructure has
significant geographic clustering and hierarchical
organization~\cite{Newmann2006,InternetGeoEconomy2005},
and second, the major fraction of the acquaintances in online social
networking communities are dictated by geographic
proximity~\cite{SocialGeography2005}. Thus, if we expect that social
interaction will dominate the cyber-traffic in near future, from
sending messages, emails and blogs to sharing videos, photos and
musics, geography can be used as the basis for overlay routing
structure that would provide the desired locality properties.

The main contributions of this paper are the arguments in favor of our
positions in two related issues -- whether the overlay support for
content transport should be treated separately from the content
lookup, and whether the off-the-shelf geographic coordinates can be
used for constructing the location-based routing overlay. In the line
of our arguments, we have outlined the design principles for the
transport overlay and presented an overlay design based on adaptive
hierarchical partitioning of the geographical space. 

In Section~\ref{sec:separation}, the history of peer-to-peer research
is analyzed to demonstrate the need for separation of the overlay
supports for lookup and transport. Section~\ref{sec:objectives}
outlines the design goals of an overlay structure for large-scale
content transport. Section~\ref{sec:geography} places the arguments
for using geographic coordinates in a location-based
overlay. Section~\ref{sec:structure} gives a brief description of the
proposed overlay structure and its routing techniques. How the
structure would be useful in content-sharing applications is explained
in Section~\ref{sec:app}.

%% file: separation.tex
\section{Development of P2P Overlays in Retrospect: Separation of
  Lookup and Transport in Content Sharing}
\label{sec:separation}
Although many different applications have been cited that could use
peer-to-peer overlays~\cite{CommonAPI2003}, the major driving force
behind the design of almost all the overlays is the single most
popular application of content sharing -- that allows a huge number of
Internet-connected end-hosts to participate in sharing of large data
contents like software packages, media files or live audio/video
streams.

Two related but subtly distinct necessities of this content-sharing
application influenced the design of the peer-to-peer overlay
networks.  The first need was indexing and search -- how to quickly
find the physical location of a specific content when a description or
a name is given. The second need was data transport -- how the content
can be efficiently transported when a large number of hosts show
interest in the same content, either at the same time, or
asynchronously over a prolonged period of time.

Several generic overlays attempted to provide both the lookup and
transport features using the same message routing
infrastructure~\cite{OpenDHT2005, Scribe2002}.  Key-based routing has
been proposed as the standard service interface that can be used to
derive all the necessary functions for the content sharing
application~\cite{CommonAPI2003}. It may be observed that the initial
designs of the overlay structures were driven by the goal of efficient
lookup. The necessity of a transport structure for concurrent or
nearly concurrent transportation of content to multiple receivers came
as a secondary thought. As such, the designers of the lookup overlays
attempted to use the same infrastructure they created for lookup for
the purpose of content routing~\cite{Scribe2002}.

However, it has later been understood that the optimization objective
of the transport overlay is grossly different from that of the lookup
overlay~\cite{Rodrigues2008, Margasinski2008}. The lookup needs fast
response while transportation needs efficient use of underlying
network links. At some point, it has also been argued that, based on
the current hardware capacities and the possible scale of the overlay
networks in foreseeable future, the fancy designs of multi-hop
structured routing overlays for the lookup service are unnecessary
complexities~\cite{Rodrigues2004}. Although there have been attempts
to propose some hybrid infrastructures~\cite{Rodrigues2008} that cover
optimization objectives for both lookup and transport, we argue that
they are fundamentally different, and hence it is beneficial to
separate these features from the ground level of the architecture.

A common design decision taken by the designers of the overlay
structures, primarily designed for the lookup service, is the randomly
assigned flat identifiers for the hosts, while the overlay structures
are defined in terms of the numeric properties of the identifiers.
The arguments placed in favor of such randomness include load
balancing, placing of replicas at uncorrelated hosts and anonymizing
the source of a content. While accepting that all these features are
necessary for content sharing, researchers have argued that randomized
placement of hosts in the overlay structure is not the only way, nor
the best way, of achieving
them~\cite{CMUReport2003}. Moreover, our position is that
emphasis on such features should not preclude the conventional wisdom
of routing locality in high-volume content transportation.

Ever since the emergence of the peer-to-peer content sharing
application, there have been growing complaints and consequent
policing from the ISPs on the traffic generated by peer-to-peer
applications. Though part of the reason of this overwhelming traffic
is the sheer volume of the contents, we believe that the on-purpose
randomized message routing topologies of most of the peer-to-peer
overlays also shares part of the blame.

%% file: objectives.tex
\section{Design Principles for the Transport Overlay}
\label{sec:objectives}
In this paper we focus on the routing overlay that is primarily used
for transport of bulk data content. The Internet Protocol is
sufficiently optimized for carrying data packets between two endpoints
in the network. However, in the large scale content sharing
application that dominates the peer-to-peer world, the same content is
transported to a large number of end-points, either at the same time
or asynchronously over an extended period of time. Thus, either in
strong or in loose sense, the necessity of these content sharing
application is an overlay that can support construction of efficient
multicast trees.

For bulk data transport, whether unicast or multicast, the primary
optimization goal in choosing the transport paths is the efficient use
of the network resources. Low latency path of transport is desirable
and may even be necessary in some applications, although it comes
secondary to the resource efficiency in case of bulk-transport
applications.

The scale of the systems demands achieving these objectives through
decentralized decisions of routing. The problem has been thoroughly
studied in the realm of IP networks. It is understood that if some
simple principles are followed in local decisions, the desired global
properties emerge.

One such well-known principle is the principle of locality, which
requires that the transport path between two endpoints of the same
local region should remain within the
region~\cite{LocalityPrinciple2005}. This discourages the traffic to
take arbitrary detours causing unnecessary burden on the global
network. The same principle also yields low latency and high
reliability paths.

Another locality property that results in efficient resource usage in
multicasting is the path-convergence property, which states that paths
from a single source to multiple destinations in one locality should
have significant portion of the path shared. The smaller the area of
the locality, the larger should be the common segment. Intuitively,
this can be attained, if the localities are hierarchically divided,
and the traffic follows a direction towards destination, gradually
resolving the destination at a deeper level of the hierarchy. Such
directional routing with hierarchical resolution will be explained in
further details in Section~\ref{sec:structure}.
 
There are other design objectives that are common in all peer-to-peer
overlays, to account for the sheer scale and the dynamics of overlay
membership. The overlay structure should be adaptive and should easily
accommodate growth and shrinkage of the membership pool. The overhead
for managing the structure must be low.

%% file: geography.tex
%%% Local Variables: 
%%% mode: latex
%%% TeX-master: "paper"
%%% End: 
\section{Location Awareness: can Geography Help?}
\label{sec:geography}
As we understand from the discussion on the overlay design principles
in Section~\ref{sec:objectives}, the overlay must take into account
the physical location of the hosts and the network links with respect
to each other while routing traffic. Indeed, several structured
overlays have been designed that base their routing decisions on
location. They differ in the way they represent and utilize the
location information.

Pietzuch et al. in~\cite{Pietzuch2006} classify location-based
overlays in two classes -- proactive and reactive ones. The reactive
location based overlays, such as Meridian~\cite{Meridian2005}, take
explicit measurement of location immediately before taking each
routing decision. Such measurement provides fresh and more correct
information but the overhead is large when the system is loaded. The
proactive location based overlays use some background mechanism to
measure the relative locations of nodes and use that information for
routing decisions. Here the overhead depends on the dynamics of
overlay membership rather than system usage. The location information
is represented either implicitly in the choice of the overlay
neighbors~\cite{Tulip2005} or explicitly by placing nodes in a virtual
coordinate space~\cite{Mithos2003}.

Positioning the Internet hosts in a virtual coordinate space, called
{\em Network Coordinates}, has been an active research issue for
several years, due to the perceived usefulness of such coordinates in
solving several distributed system problems such as resource
discovery, replica placement and efficient routing. The usual choice
of the coordinate space is a multidimensional Cartesian space, and
research shows that the Internet hosts can be mapped into a
low-dimensional Cartesian space with acceptable
accuracy~\cite{Pietzuch2005}. Notable projects that approach such
mappings include GNP~\cite{GNP2002} and
Vivaldi~\cite{Vivaldi2004}. Nevertheless, high background overhead
inhibits the acceptability of network coordinates for many useful
applications.

Here we argue that, for the purpose of overlay routing adhering to the
locality principle, two dimensional geographic coordinates (latitude,
longitude) of the hosts would provide sufficient location information.
At present, there are a number of comprehensive
databases~\cite{MaxMind,GeoBytes,IP2Location} that resolve the
geographic coordinates of Internet hosts, usually at the resolution of
the point-of-presence of the ISP. Since the geographic location of
hosts at this resolution is relatively stable within a session, the
information can be obtained by off-the-shelf database lookup,
eliminating any measurement overhead.

Studies on spatial characteristics of the Internet infrastructure show
that Internet, like many other networks, have strong geographic
clustering, following the geographic distribution of its
users~\cite{InternetGeoEconomy2005,Newmann2006}. Also, like
hierarchical division of geographic locations made for political and
administrative purposes, Internet backbone can also be roughly
organized in different tiers, that serve interconnection for locations
at different levels such as continent, country, state and
city~\cite{InternetHierarchy2002}. It is true that there are many
instances of different ISPs in the same geographic location, and the
ISP's preferred route for geographically local traffic between two
ISPs may not exactly follow the locality principle at fine
resolution. Nevertheless, transporting the traffic between points of
the same geographic location locally is arguably beneficial for the
globally optimal use of Internet resources. The growth of caching
proxy networks like Akamai to contain the local web traffic locally
also supports the argument.

Indeed, before the advent of the peer-to-peer overlays, there has been
attempts to introduce geography-directed routing in the Internet
infrastructure~\cite{GPSRouting1999}. Although the technique faced
deployment hurdles in the rigid infrastructure, the applications it
envisioned only became more relevant in present days. Besides
resource-efficient routing, the applications include location-based
information search, finding nearest services and broadcasting messages
in a geographic community.

Recent growth of social networking platforms suggests that social
acquaintance of network users will dictate the direction of majority
of content transport in the Internet in the near
future. Interestingly, a recent study shows that more than two thirds
of the acquaintances in on-line social networking communities are
defined by geographic locality~\cite{SocialGeography2005}. If the
general purpose routing overlay is to be used for future
implementations of these social applications, the finding supports the
argument that the overlay should be carefully optimized for
geographically localized traffic.

%% file: structure.tex
%%% Local Variables: 
%%% mode: latex
%%% TeX-master: "paper"
%%% End: 
\section{Structure of the Overlay Interconnection}
\label{sec:structure}
In the previous sections, we argued in favor of a location-based
routing overlay for peer-to-peer applications and that geographic
coordinates can serve as the basis of such location-based overlays.
In this section, we present an overlay interconnection structure based
on hierarchical partitioning of the geographic space, where traffic is
routed towards the geographic location of the destination,
successively resolving the destination at a deeper level of the
hierarchy.

\begin{figure*}[htb]
  \centering
  \setcounter{subfigure}{0}
  \subfigure[Zoning of geographic clusters. 
            Routing paths show locality ($h\rightarrow k$) 
            and path convergence ($h\rightarrow(q,r)$)]{
    \label{fig:routing_range}
    \includegraphics[scale=0.85]{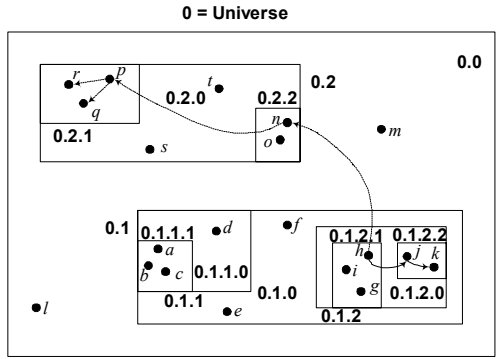}
  }
  \hspace{1cm}
  \subfigure[Zoning hierarchy and overlay neighborhood of peer {\em h}]{
    \label{fig:zone_tree}
    \includegraphics[scale=0.85]{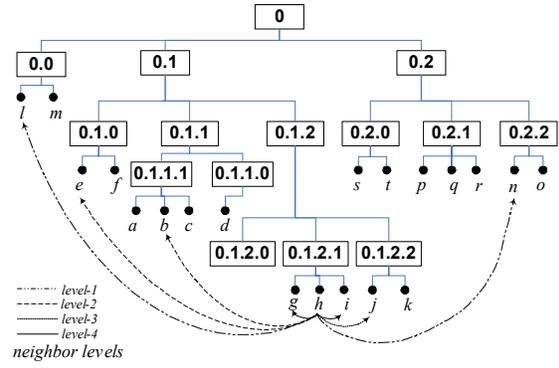}
  }
  \caption{Routing overlay by hierarchical zoning}
\end{figure*}

\subsection{Structure}
The universe (earth surface) is hierarchically divided into zones,
sub-zones, sub-sub-zones and so on. A zone is divided into
non-overlapping sub-zones and the higher-level zone completely covers
all the areas of its sub-zones. The shape of the zones need to be
amenable to concise memory representation and also to easy computation
of whether a point belongs to a zone or not. A simple shape such as an
axis-parallel rectangle may be used as zones. At the leaf level of the
hierarchy are the zones that are not divided any further (denoted {\em
  leaf zone}). Each individual overlay node or {\em peer} belongs to a
leaf zone at its deepest level, to successively larger zones at higher
levels, and to the zone covering the universe at the top
level. Figure~\ref{fig:routing_range} illustrates an example division
of the universe and the corresponding tree representation is shown in
Figure~\ref{fig:zone_tree}.

A {\em routing table} in each peer stores the overlay neighbors of the
peer. To be able to route messages towards a destination by
successively resolving the zones at finer grain, a peer need to know
at least one peer in all the sibling zones at every level of the
hierarchy. At the deepest level, the peer knows all other peers within
its own leaf zone. The routing table may be organized in rows, each
row storing the pointers to the siblings at a different level.  For
each pointer, the IP address of the target peer and the boundary
definition of the corresponding sibling zone is stored. Also, the
boundary definition of the self-zone at every level is stored at the
corresponding row. Additionally, peers and zones can be uniquely
identified globally, using a hierarchical name that concatenates the
identifications of the zones at successive levels of the hierarchy (as
shown in the figure). Such names (denoted as {\em overlay identifier})
can be stored in each entry of the routing table, besides the zone
boundary definition. The overlay neighborhood of a peer is illustrated
in Figure~\ref{fig:zone_tree}.

\begin{algorithm}[h]
\caption{RouteToAllPeers($msg$, $area$, $level$)}
\label{alg:route_all}
\begin{algorithmic}[1]
  \IF {$peer.coordinate$ falls in $area$}
     \STATE{Deliver $msg$ to $peer$} \label{algRQ:selfCheck}
  \ENDIF 
  \IF {$level \leq$ deepest level $d$}
    \FOR {Each entry $e$ in row $d$ of the routing table}\label{algRQ:selfzoneloop_start}
      \IF {$e.coordinate$ falls in $area$} \label{algRQ:cond_fallsin}
	 \STATE{Send new $RouteToAllPeers(area, d+1, msg)$ to $e.IPAddress$} 
      \ENDIF 
    \ENDFOR \label{algRQ:selfzoneloop_end}
  \ENDIF 
  \FOR {$r=d-1$ down to $level$}
  \label{algRQ:rowloop_start}
    \FOR{each entry $e$ in row $r$ of the routing table, except for the
    one denoting self zone}
      \IF{$e.zone\_boundary$ intersects $area$}\label{algRQ:cond_intersect}
        \STATE{Send new $RouteToAllPeers(area, r+1, msg)$ to $e.IPAddress$}
      \ENDIF
    \ENDFOR
 \ENDFOR \label{algRQ:rowloop_end}
\end{algorithmic}
\end{algorithm}

The beauty of the overlay structure lies in its flexibility to grow
and retract with the membership dynamics, and its ability to manage
this in a completely decentralized way. When the number of peers in a
leaf zone grows beyond a threshold, new sub-zones are created by
dividing the zone according to geographical clusters of peers. Note
that all peers in a leaf zone are neighbors to each other in the
overlay. So any peer, knowing coordinates of all other peers in the
same leaf zone can perform the partitioning and inform all others of
the new boundaries and identifiers. Similarly, when it is discovered
that the number of peers in a leaf zone is below a threshold, the leaf
zone can initiate a merge with one of its siblings. When new sub-zones
are created based on geographic clustering, an area of the previous
leaf zone that does not belong to any of the clusters, is also
considered a sub-zone and is denoted as remainder-leaf-zone. The
remainder-leaf-zone always serves as a suitable merger siblings for
the other leaf zones. Details of the adaptation techniques in response
to membership dynamics can be found in~\cite{spatialSearch}.

\subsection{Routing}
The overlay is able to route messages towards a geographic
location. The target may be all or any peers in the specified area, or
the nearest or a nearby peer of a specified point. Messages may also
be routed to a particular peer specified by the overlay identifier.

To forward a message targeted to an area, a peer uses the {\em
RoutToAllPeers} method defined in Algorithm~\ref{alg:route_all}. A
peer forwards the message to its contacts in all sibling zones at all
levels of the routing table, whose zone-area intersects with the
target area (Lines~\ref{algRQ:rowloop_start}-\ref{algRQ:rowloop_end})
and to all peers within the leaf level self-zone that fall in the
target area
(Lines~\ref{algRQ:selfzoneloop_start}-\ref{algRQ:selfzoneloop_end}). To
remember the levels of hierarchy already resolved, the $level$
parameter is used, which is set to $1$ at the peer that initiates the
routing. 

Routing a message to a peer at or near a specified point can be
performed by the same algorithm with minor modifications. The
algorithm will have a $point$ as the third parameter instead of an
$area$. The condition in Line~\ref{algRQ:cond_intersect} will check if
the $point$ falls in the zone and the loop in
Lines~\ref{algRQ:rowloop_start}-\ref{algRQ:rowloop_end} will terminate
as soon as a match is found. The loop in
Lines~\ref{algRQ:selfzoneloop_start}-\ref{algRQ:selfzoneloop_end} will
forward the message to the peer in the self zone closest from the
target.

If precisely the peer nearest to a specified point is sought, it can
be done by first reaching the peer that is closest to the peer within
the zone that holds the point, and then sending a query message
towards a circular area with the target point at its center and the
current peer at the perimeter, to figure out if any other peer closer
to the target exists.

By construction, it is observable that the overlay routing adheres to
both the locality and the path convergence principles outlined in
Section~\ref{sec:objectives}. An illustration of the routing paths in 
Figure~\ref{fig:routing_range} demonstrates both the properties.

%% file: app.tex
%%% Local Variables: 
%%% mode: latex
%%% TeX-master: "paper"
%%% End: 
\section{Application of the Routing Overlay}
\label{sec:app}
Focusing back to the origin of peer-to-peer research, i.e.\ the large
scale content sharing application, we argued in
Section~\ref{sec:separation} in favor of separation of its search and
transport services. The generalized routing overlay is designed having
in mind mainly the requirements for the high-volume data transport to
a large number of recipients. Here we explain how the overlay may be
used for multicasting and proactive caching.

A request for the content is sent to the source directly. Because there
is a separate service for lookup, the IP address of the source of a
desired content can be known from there.  The routing overlay is used
for efficient delivery of the content once the request is
received. Knowing the overlay identifier of the requester, the source
is able to explore the overlay route towards the requester. Due to its
locality and path convergence propoerties, this route can be used to
create an efficient delivery tree for live streaming content, when
there are many requests for the same content~\cite{CliqueStream}.

For non-live contents that are shared among many users during an
extended period of time, replicas of the fragments of the content can
be stored along the overlay route. This helps to serve multiple
requests originated in the same geographic locality from the nearest
replica without causing unnecessary traffic burden on the long
distance links.

Besides multicast transport, the routing overlay can be used for
several geography related applications, such as, finding the nearest
service or looking up services in a geographic range from a location,
or broadcasting some message or command to the hosts in a specific
geographic region~\cite{spatialSearch}.

%% file: conclusion.tex
%%% Local Variables: 
%%% mode: latex
%%% TeX-master: "paper"
%%% End: 
\section{Conclusion}
In this paper, we raised the issue of separating the transport from
lookup in the design of overlays for peer-to-peer content sharing, and
explained how geographic coordinates of the hosts can be used to
create a location-based overlay that yields efficiency of resource
usage in bulk transport. Whether such generalized overlay will be
useful for other unforeseen applications, can only be examined in the 
future.